\documentclass[english,journal]{IEEEtran}
\usepackage[T1]{fontenc}
\usepackage[latin9]{inputenc}
\usepackage{amsmath}
\usepackage{amssymb}
\usepackage{amsthm,bm}
\usepackage{url}

\makeatletter
\theoremstyle{plain}

\theoremstyle{remark}

\makeatother

\usepackage{babel,verbatim,balance}
\usepackage[rgb]{xcolor}
\usepackage{graphics, graphicx}
\usepackage{acronym}
\usepackage{upgreek}
\usepackage{setspace}

\acrodef{bs}[BS]{base station}
\acrodef{csi}[CSI]{channel state information}
\acrodef{tdd}[TDD]{time division duplexing}
\acrodef{ris}[RIS]{reconfigurable intelligent surface}
\acrodef{rss}[RSS]{received signal strength}
\acrodef{toa}[ToA]{time of arrival}
\acrodef{poa}[PoA]{phase of arrival}
\acrodef{aoa}[AoA]{angle of arrival}
\acrodef{aod}[AoD]{angle of departure}
\acrodef{ula}[ULA]{uniform linear array}
\acrodef{snr}[SNR]{signal-to-noise ratio}
\acrodef{gdop}[GDOP]{geometric dilution of precision}
\acrodef{em}[EM]{electro-magnetic}
\acrodef{fim}[FIM]{Fisher information matrix}
\acrodef{los}[LoS]{line-of-sight}
\acrodef{nlos}[NLoS]{non-line-of-sight}
\acrodef{bs}[BS]{base station}
\acrodef{peb}[PEB]{position error bound}
\acrodef{uwb}[UWB]{ultra wideband}
\acrodef{slam}[SLAM]{simultaneous localization and mapping}
\acrodef{psd}[PSD]{power spectral density}
\acrodef{ofdm}[OFDM]{orthogonal frequency division multiplexing}
\acrodef{rt-tof}[RT-TOF]{round-trip-time-of-flight}
\acrodef{tdoa}[TDoA]{time-difference-of-arrival}
\acrodef{coa}[PoA]{phase of arrival}
\acrodef{mse}[MSE]{mean squared error}
\acrodef{slam}[SLAM]{simultaneous localization and mapping}
\acrodef{ue}[UE]{user equipment}

\newcommand{\rev}[1]{{\textcolor{black}{#1}}}
\newcommand{\revB}[1]{{\textcolor{black}{#1}}}

\providecommand{\remarkname}{Remark}
\providecommand{\theoremname}{Theorem}
\begin{document}

\title{Radio Localization and Mapping with Reconfigurable Intelligent Surfaces}

\author{Henk Wymeersch\IEEEauthorrefmark{1}, Jiguang He\IEEEauthorrefmark{2}, Beno\^{i}t Denis\IEEEauthorrefmark{3}, Antonio Clemente\IEEEauthorrefmark{3}, Markku Juntti\IEEEauthorrefmark{2}\\
\IEEEauthorrefmark{1}Chalmers University of Technology, 
\IEEEauthorrefmark{2}University of Oulu, 
\IEEEauthorrefmark{3}CEA-Leti
\thanks{Henk Wymeersch was supported by the Swedish Research Council under grant 2018-03701. The work by Jiguang He and Markku Juntti has been supported in part by the IIoT Connectivity for Mechanical Systems (ICONICAL) project and 6Genesis Flagship (grant 318927), both funded by the Academy of Finland.}}

\maketitle
\begin{abstract}
\rev{5G radio at millimeter wave (mmWave) and beyond 5G concepts at 0.1--1 THz can exploit angle and delay measurements for localization, by the virtue of increased bandwidth and large antenna arrays but are limited in terms of blockage caused by obstacles. 
Reconfigurable intelligent surfaces
(RISs) are seen as a transformative technology that can control the physical propagation environment in which they are embedded by passively reflecting EM waves in preferred directions. 
Whereas such RISs have been mainly intended for communication purposes, RISs can have great benefits in terms of performance, energy consumption, and cost for localization and mapping. These benefits as well as associated challenges are the main topics of this paper. }

\end{abstract}

\section*{Introduction}

The interaction between the digital and physical world relies on high-definition situational awareness, i.e., the ability of a device to determine its own location, as well as the location of objects and other devices in the operating environment. Applications include automated vehicles and robots in general, as well as healthcare, {highly immersive virtual and augmented reality, or new  human-to-machine interfaces}. 
%
Situational awareness can be achieved by a variety of technologies, depending on the application and requirements, including lidars, inertial measurement units, or cameras, but also radio-based technologies, such as satellite positioning, radar, \ac{uwb}, cellular or WiFi.
Radio-based technologies are attractive as they can have a dual communication and sensing functionalities and are often less susceptible to environmental factors such as poor lighting. 
Since 4G, dedicated localization reference signals have been considered as part of communications system design and standardization. These can enable location accuracy levels on the order of 10 m. With 5G, the use of larger bandwidths and higher carrier frequencies in combination with antenna arrays at the {\ac{ue} and \ac{bs}} is expected to further improve the location accuracy to around 1 m. Within Beyond 5G systems, the trend is to operate at much higher frequencies (above 30 GHz, possibly up to 1 THz) so as to benefit from large available bandwidths and thus achieve even better localization accuracy. 

Propagation at high carrier frequencies 
suffers from obstructions due to objects blocking \ac{los} path between the transmitter and the receiver. The reliance on the \ac{los} path can be reduced through multipath-aided localization by exploiting either a prior map information   or through joint localization and mapping
\cite{leitinger2018belief}. Therein,  the locations of objects in the environment (surfaces and scatter points) are determined simultaneously with the user's location, a process called  radio-based \ac{slam}. 
%
Even if these solutions make use of the multipath channel as a constructive source of information in the localization problem geometry, the related electromagnetic interactions (induced by the physical environment) still remain uncontrolled and as such, largely sub-optimal from a localization perspective. 

 \begin{figure}
 \centering
 \includegraphics[width=0.8\linewidth]{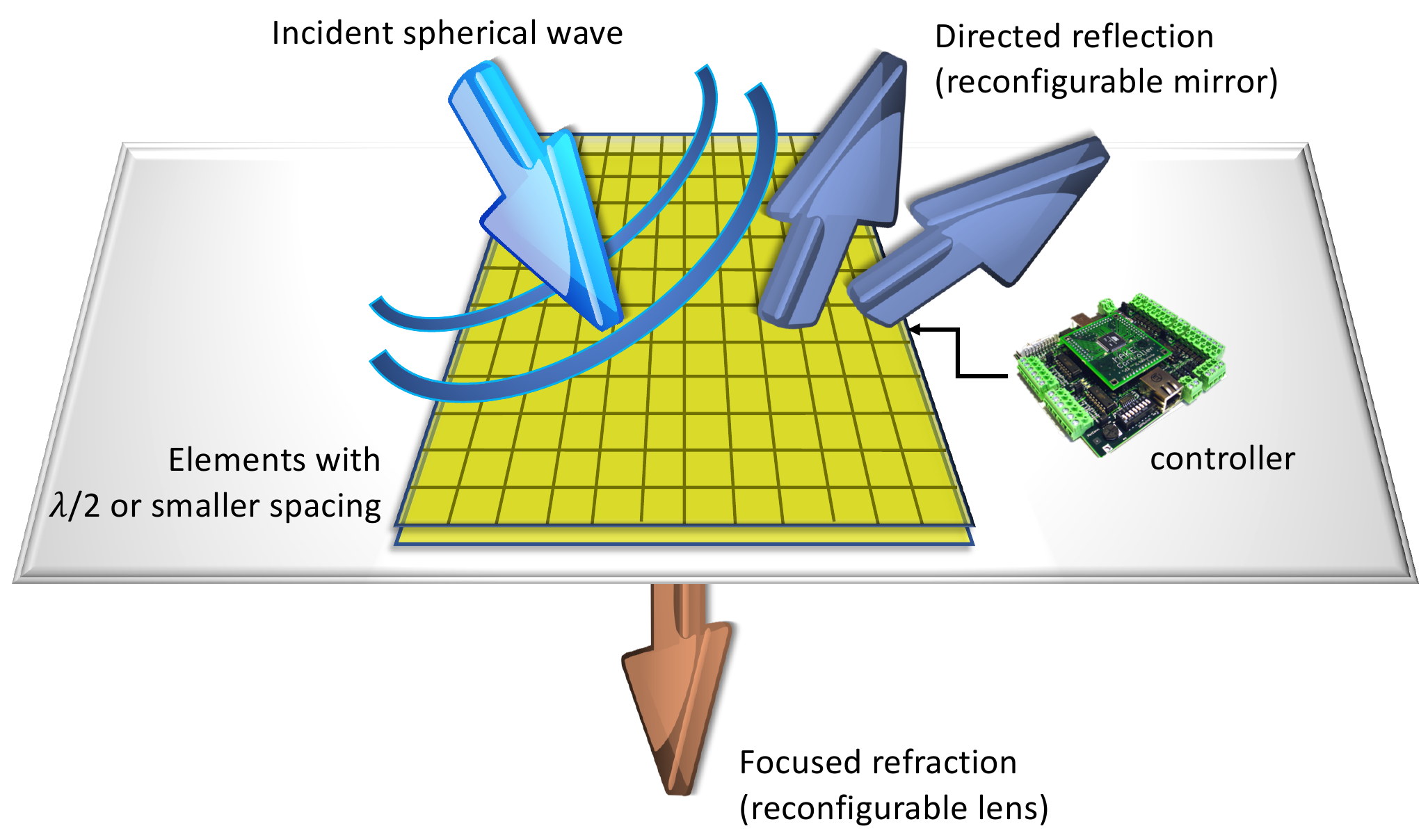}
 \caption{\revB{Example of a RIS, where a controller adjusts individual elements. Depending on the technology, the RIS can change the direction of reflections or refract the signal, similar to a lens. }}
 \label{fig:RISmodel}
 \end{figure}

 \begin{figure*}
 \centering
 \includegraphics[width=0.9\textwidth]{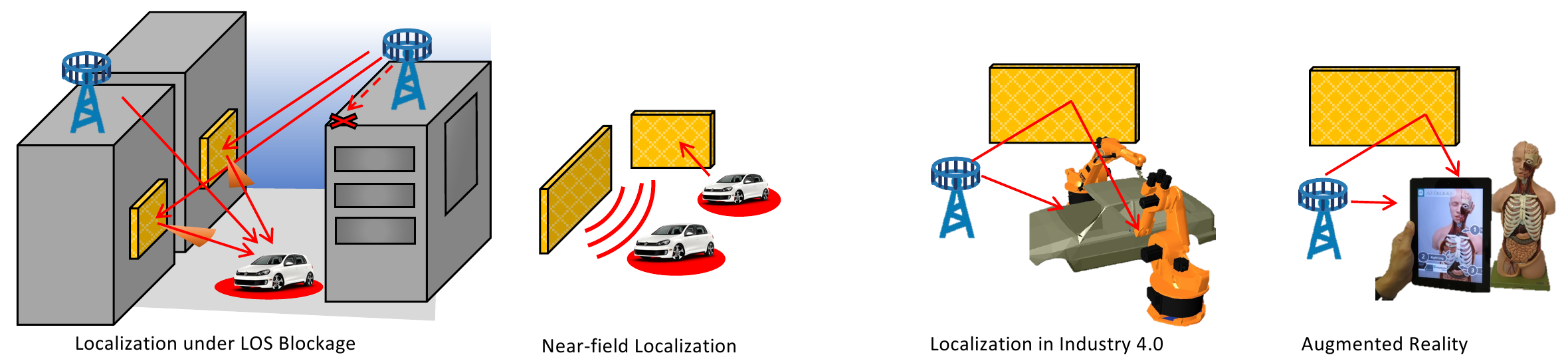}
 \caption{Application examples of RIS-based localization and mapping services \rev{ (from left to right): (i) \ac{los} blockages can be circumvented  to improve localization accuracy and continuity; (ii) wavefront curvature in the near-field of a large RIS can be exploited to solve for nuisance parameters (e.g., clock biases); (iii) by creating strong and consistent multipath, RISs can support localization in very harsh indoor environment, dynamically accounting for object movements; (iv) new delay-sensitive, ultra-accurate applications will be supported by the fact that RISs do not introduce processing delays.}}
 \label{fig:applications}
 \end{figure*}
Reconfigurable intelligent surfaces (RISs) \acronymused{ris} represent a breakthrough technology
whereby surfaces are endowed with the capability to actively modify the impinging electromagnetic wave \cite{2019Basar}\revB{, as visualized in Figure \ref{fig:RISmodel}. A RIS can be implemented using a variety of technologies as discussed below and can provide significant benefits in terms of communication  by guaranteeing coverage when the \ac{los} is blocked. A RIS can operate as a reconfigurable mirror or as a reconfigurable lens (see Figure \ref{fig:RISmodel}). The RIS is controlled by a local control unit that adjusts the phase profile or current distribution. Based on these fundamental operating modes, a RIS can act as transmitter \cite{2019Basar}, receiver \cite{abdelrahman_taha_deep_2019} or as 
%
%
%
%
an 
anomalous reflector, where the direction of the reflected wave is no longer specular according to natural reflection laws but steerable
\cite{bjornson_intelligent_2019,Wu2019intelligent}. The RIS concept can be applied at different wavelengths, ranging from low sub-6GHz bands, where the technology is well understood and commercial systems are available, to 28 GHz mmWave bands, where RISs can provide significant benefits in terms of coverage but where the technology is less mature. Finally, in the 0.1--1 THz regime, severe path loss, higher susceptibility to blockages, atmospheric absorption, and rain attenuation as well as significant hardware limitations make RIS design challenging but can also lead to high performance gains.}


\revB{The aforementioned properties and their close relation to the environment geometry make RIS attractive for localization and mapping. 
The potential of RIS for localization has received only limited coverage in the literature,  including preliminary studies where the RIS operates in receive mode as a lens \cite{hu_beyond_2018} and in reflection mode \cite{he2019adaptive}. }Hence, it is timely to delve deeper into the potential of RIS for localization and mapping, as well as the main research questions that we should address in the coming years. Possible applications of RIS for localization are visualized in Figure \ref{fig:applications}.

The aim of this paper is to \rev{take a broader view than the technical contributions in \cite{hu_beyond_2018,he2019adaptive} by describing}  the core technical challenges of applying RISs to localization and mapping, {along with a preliminary system vision, results and solutions recently put forward on related topics.}


\section*{Radio localization and Mapping}

\subsection*{Basic Principles}


Any radio localization and mapping system comprises three essential parts: measurements, a reference system, and the inference algorithms. 

\subsubsection*{Measurements}
    The measurements are derived from the radio signal between a transmitter and a receiver. They can typically be obtained directly from the channel estimation routine used for communication. 
    \revB{Common {location-dependent metrics} are based on received signal strength, \ac{toa}, \ac{poa}, \ac{aoa} and \ac{aod}, and Doppler measurements.} 
    %
    %
    Measurements can be characterized by their \emph{resolution} and \emph{accuracy}. The resolution refers to the ability to distinguish two signals based on their measurements and depends
    on the signal bandwidth and duration, carrier frequency, the number of antennas, and coherent integration time. The accuracy refers to the extent to which we can determine the parameter of interest. It depends also on the \ac{snr}, as well as on the detailed properties of the signal waveform such as the time-frequency and spatial power allocation. 
    
    \subsubsection*{Reference System}
    \revB{All the measurements are taken in a certain frame of reference, {e.g.,} that of the receiver. References, sometimes called anchor points, have known states. 
    There may be multiple position references, as in cellular localization or satellite positioning, which may in turn place requirements in terms of synchronization, array calibration, as well as dedicated control signals. The geometric placement of the reference plays an important role in the accuracy of a localization system, an effect commonly measured through the \ac{gdop}. }

    \subsubsection*{Localization and Mapping Algorithms}
     A main distinction between a communication and a localization algorithm is how the multipath is treated. In communication, multipath is used to provide diversity or spatial multiplexing, thus, decreasing the error rate or increasing the data rate. In localization, only the \ac{los} has traditionally been used, as the measurements associated with that path could directly be related to the location of the user. Modern approaches also exploit measurements from \ac{nlos} paths, corresponding to scattered or reflected signal components~\cite{leitinger2018belief}. 
     %
     A critical component in \ac{slam} is the association of measurements to their sources, where a source can be a transmitter or a fixed object in the environment, or clutter.

In the design aspects of the measurements, reference system, and algorithms,  fundamental performance bounds can play an important role. They allow us  to assess the localization potential of signals or reference systems, {guide the development and benchmarking of algorithms, or can even be used as approximated performance indicators or real-time optimization/selection criteria.}

\subsection*{Localization and Mapping with RIS}

The inclusion of RIS affects the three above-mentioned parts of radio localization. 
The measurements are in general tuples of \ac{toa}, \ac{poa}, \ac{aod}, \ac{aoa}, and the Doppler shift. The relation between the measurements depends on the underlying channel model, which 
is largely geometric: each path corresponds to a cluster of rays, depending on the \ac{em} properties of the objects. In other words, the locations and \ac{em} properties of the environment impose a mapping from position space to measurement space. \rev{Whether this mapping is resolvable depends on the available bandwidth and number of antennas. While RIS can be used at sub-6 GHz, the 
larger bandwidths at frequencies beyond 28 GHz, combined with more dense packing of RIS elements are particularly conducive to localization and mapping.}
%
The references include the BS and RIS, which can reasonably be assumed to have a pre-programmed known location and orientation in a common coordinate system, while users and passive object have unknown or partially known location and orientation information. The signal from the BS is to a large extent controllable in the time, frequency, and spatial domains. Therefore, it can be optimized in terms of power allocation and beamforming to maximize the accuracy of the measurements. The signals from the RIS can be shaped by the RIS controller, in order to further improve accuracy, when the RIS is acting as a transmitter or a reflector~\cite{hu_beyond_2018, he2019adaptive}. The design may however be less flexible than the signal from a conventional BS, {for obvious  power and complexity considerations}.
In terms of inference algorithms, RIS-based \ac{slam} 
should harness the flexibility of the BS signals and RIS controllability, to improve not only localization and mapping coverage, but also accuracy. 



\section*{Challenges and Opportunities}
When a RIS used as a reflector, it could be interpreted in two different ways: as part of the passive environment, acting like any scatterer or reflector, or alternatively as part of the infrastructure, playing a similar role as a global reference or anchor point. These two views lead to fundamental challenges and opportunities \rev{(in terms of applications and research directions)} in incorporating RIS in radio localization and mapping, as highlighted below. Many of these challenges are interrelated, but are presented as separate for reasons of clarity: RIS and channel modeling, near-field propagation, channel estimation, {system} architecture and signaling, RIS control, waveform design, and SLAM methods.

\subsection*{RIS Modeling and Channel Modeling} 
\subsubsection*{Challenge} 
There are several different {antenna technologies} and terminologies for RIS, including {classical phased arrays}, reflectarrays {\cite{Hum2014}, transmit arrays \cite{DiPalma2017}}, smart, programmable or software defined metasurfaces {\cite{Martini2019}}, large intelligent surfaces (LISs), {etc}. Making their usage truly ubiquitous, programmable wireless environments could be created \cite{Liaskos18}. Proper models of their functionality or how they interact with \ac{em} waves still represent an active area of research.  {As in the case of the beamforming, RISs could be implemented as full-digital, hybrid or analog architectures with both amplitude and phase or phase-only control. Quasi-continuous phase quantization could be selected as a function of the required complexity and power consumption specifications.} 

{In the RIS model,} the radiation pattern in azimuth and elevation should account for coupling of the RIS elements{, which are typically located on a regular or triangular lattice with an inter-element distance ranging between one-tenth and one-half wavelength.}
Impedance matching, reflection and refraction losses can affect RIS performance. {In the case of reflect and transmit arrays, for example, the scattering properties of the elements should be included in the model. The impact of the oblique incidence on the element performance is also an important parameter. {More generally, this RIS model should be defined according to the \ac{em} properties of the chosen underlying technology (e.g., specific \ac{em} synthesis tools are needed to calculate the impedance modulation in case of metasurfaces). RIS geometry and periodicity, which impact the mutual coupling between its constituting elements, should also be taken into account. Finally, the method and electronics used to control the RIS beam (e.g., single frequency phase-shift, time delays, quasi continuous phase vs.~quantized phase) should be properly developed, while considering related hardware impairments (e.g., specific models for phase-shifters and other building tunable devices, including RF losses and limited resolution, active element performances)}}.
The model of the radio channel to and from a RIS, including the beam shape of signals, polarization effects, path loss, as well as joint angular and  delay spread and how to control these require significant research efforts. Moreover, the interaction with new BS technologies and radio stripes is poorly understood. 
\rev{Hardware impairments will be more pronounced the higher the carrier frequency, which in turn impacts the amount of flexibility and control needed. }

\begin{figure}
 \centering
 \includegraphics[width=1\columnwidth]{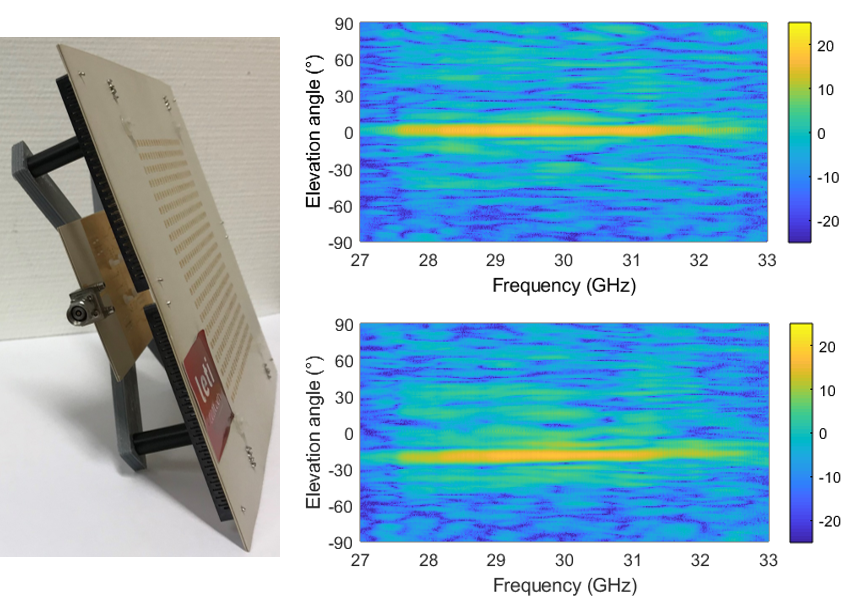}
 \caption{{Ka-band RIS based on electronically steerable transmitarray architecture. (Left) $20\times20$-element RIS with 800 p-i-n diodes and a substrate integrated wave-guide (SIW) spatial feed \cite{DiPalma2017}. (Right) Measured co-polarization beams (gain in dBi) on the $0^{\circ}$-azimuth cut-plane at broadside direction and scan angle of $-20^{\circ}$ as a function of the elevation angle and frequency.}}
 \label{fig:RIS_TA}
 \end{figure}

\revB{While more in-depth overview of different technologies can be found in the literature for reflectarrays \cite{Hum2014}, transmitarrays \cite{Reis2019}, and phased arrays \cite{PhasedArray}, a specific example of RIS based on transmitarray technology is presented in Figure \ref{fig:RIS_TA}. This antenna is composed of a controllable flat lens with $20\times20$ elements, having an aperture size of $102 \times 102~\text{mm}^2$  and a spatial feed based on a 16-element substrate integrated wave-guide (SIW) array \cite{DiPalma2017}, located at a distance of $30~\text{mm}$ from the flat lens aperture. 
A beam can be electronically controlled with 1-bit phase quantization at a commutation speed in the range $5-10~\text{ms}$. The aperture efficiency corresponds to a realized gain between $20-23$ dBi, which can be improved up to $40\%$ with 2-bit designs. 
}

\subsubsection*{Opportunities}
Determining proper models requires a combination of skills, ranging from the \ac{em} theory to circuits. 
Since there are multiple RIS technologies and a RIS can act in transmit, receive, or reflect mode, there is no one-size-fits-all model. What is common in all these models, however, is the dependence on location, orientation, and extent of the RIS, leading to 
clear opportunities to use the models for localization purposes, where each  specific model of RIS may present different opportunities to improve localization and mapping. \rev{In addition, if models are to be used for localization and mapping, they should be spatially and temporally consistent and account for the locations and orientations of all relevant objects (both passive and active). Models are likely to differ, depending on the frequency band, but multi-band operation can benefit from models that are consistent across a very wide frequency range.}

\subsection*{Near-field Propagation}

\subsubsection*{Challenge} 
Beyond the Frauhofer distance, signals are in far-field so that the plane wave assumption holds. The near-field region is proportional with the surface area of the RIS, so that a $20~\text{cm} \times 20~\text{cm}$ RIS has an 8 meter near-field region at a wavelength of 1 cm. Hence, even at moderate distances to the RIS, near-field propagation occurs, leading to wavefront curvature, which must be properly modeled and accounted for in the communication system. This affects both RIS and channel modeling as well as RIS channel estimation and control. 

\subsubsection*{Opportunities} As regards near-field propagation, the wavefront curvature (see also Figure \ref{fig:applications}) can be harnessed to reduce the need for infrastructure or synchronization. The \ac{poa} from a near-field signal provides information about both the angle and distance to the RIS, so that in combination with \ac{toa} it is possible to determine unknown clock biases. The \ac{poa} observable by an array of elements, possibly asynchronous and non-coherent to the transmitter itself, can also be exploited directly in terms of spherical wave localization. 
This exploitation requires novel dedicated signal processing methods, as well as possibly new signal designs that can maximally harness the near-field properties. The specific properties of different RISs (e.g., their size) can be used in near-field multipath-aided positioning and to simplify data association in \ac{slam}. 



\subsection*{Channel Estimation}
\subsubsection*{Challenge}
In communication, for the purpose of detection, phase adjustment, or precoding, RIS channel estimation is needed in receive, reflect, or transmit mode, respectively. 
\rev{For localization, the compound channel  needs to be estimated at the receiver side, in order to extract the \ac{aoa}, \ac{aod}, and \ac{toa} of each propagation path (or cluster), as well as their respective spreads. }
As a RIS may have limited processing capabilities and, under reflect mode, may have no or few RF chains, such channel estimation to and from the RIS is challenging \cite{abdelrahman_taha_deep_2019}.   For instance, in \cite{zheng2019intelligent}, a protocol is proposed to separately estimate the \ac{los} and {RIS} channels, by activating the RIS with different phase patterns while sending pilots leading to delays. 
Channel estimation in receive mode is arguably not well understood, with, e.g., \cite{jung2019uplink} analyzing the impact of channel estimation errors, but not proposing a channel estimation routine. 


\subsubsection*{Opportunities}
At high carrier frequencies, the channel response is sparse 
and depends mainly on the geometric configuration of UE, BS, and the environment (including the RIS). { Hence, the sparse channel properties can be leveraged in the process of channel parameter estimation by resorting to compressive sensing (CS) methods~\cite{abdelrahman_taha_deep_2019}. The estimated channel parameters in turn help to determine the user location via the 3D geometrical relationships.} Prior location information of the UE and the RIS location and orientation could be used as a proxy for \ac{csi}. In other words, the geometric information could be converted to partial \ac{csi} or to \ac{csi} statistics. For instance, the end-to-end compound channel  can be determined a priori as a function of the UE location through machine learning techniques. As the UE location is generally only statistically known, this uncertainty should be reflected in the \ac{csi} uncertainty {accordingly}. Hence, suitable Bayesian methods are needed to provide this mapping.

 \begin{figure*}
 \centering
 \includegraphics[width=0.65\linewidth]{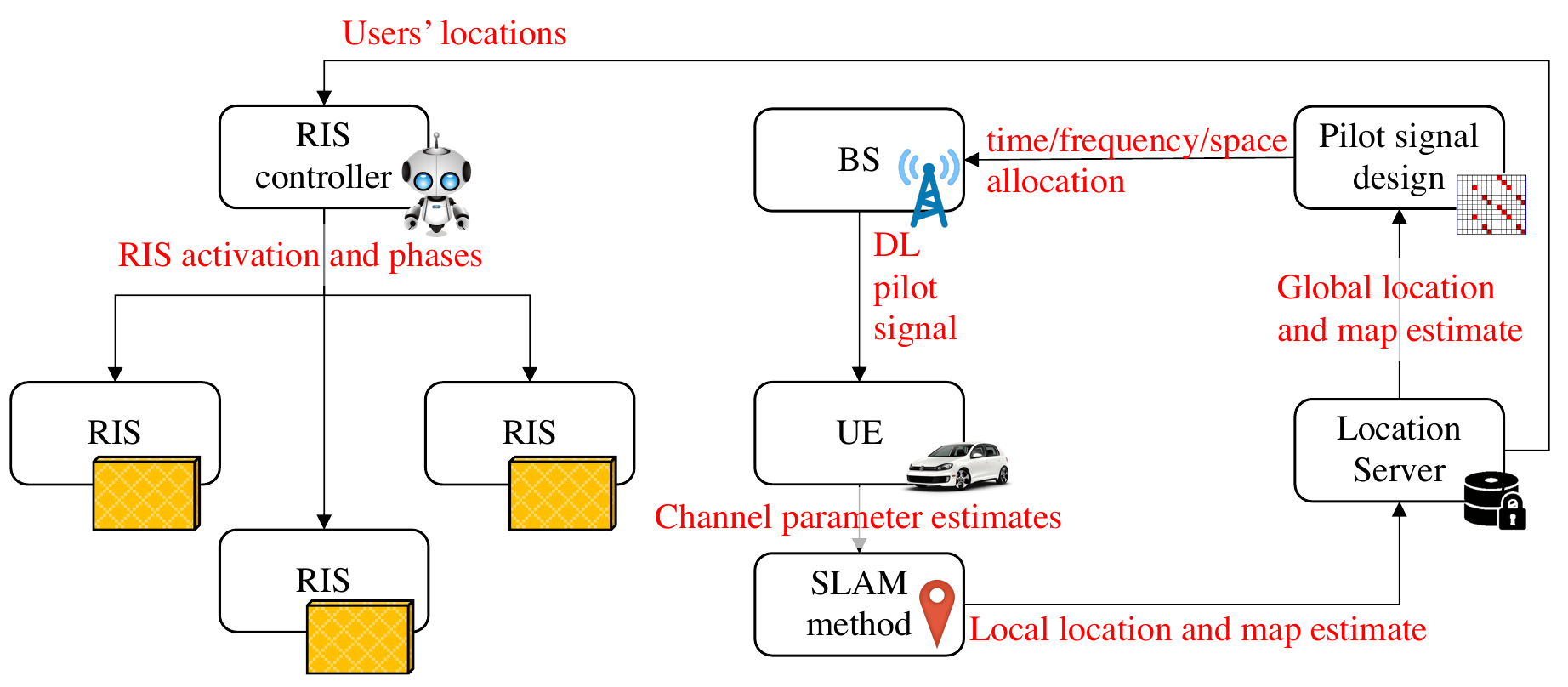}
 \caption{Possible flowchart of signaling for downlink RIS-aided localization and mapping. A priori user location information is used to determine which RIS to activate and how to set its phases. The downlink (DL) pilot signal, reflected by the RIS, is optimized given the current UE and environment conditions, and 
is used by the UE to estimate the channel parameters. These are fed to the SLAM algorithm, which determines the UE location and local map. Maps from different UEs can be fused to provide global situational awareness.}
 \label{fig:flowchart}
 \end{figure*}

\subsection*{Signaling and {System} Architecture}
\subsubsection*{Challenge} 
Localization can be performed in uplink\rev{, downlink,} or sidelink (i.e., between two \acp{ue}). Uplink localization can benefit from richer measurements and more processing power at the BS side, while downlink localization can reuse high-power downlink pilots, localize multiple users simultaneously, and requires less UE power. Sidelink signals can be used for relative localization, both in a bistatic and a monostatic configuration. No matter which architecture is chosen, control and feedback signals need to be provided among all network entities. Calibration and synchronization signals are needed for maintaining coherence among the position references. 
This can be performed over the air or via wired links between the infrastructure elements. Finally, fine a priori location and orientation information of RIS is needed to support localization. \rev{For static RIS, this can be achieved by a one-time surveying step or by the use of GPS signals when available. For mobile RIS, the architecture should support RIS tracking methods.}


\subsubsection*{Opportunities}
The design of signaling protocols and the trade-offs of uplink, downlink, and sidelink RIS-aided localization {are still unknown and remain largely unexplored in the research community}. A possible architecture with the corresponding signal flow is depicted in Figure \ref{fig:flowchart}. 
The estimated UE location information can be re-injected to refine the RIS setting and selection to further improve the next localization steps.  As RISs are expected to often operate with obstructed \ac{los}, localization and mapping methods can support communication by providing the system with prediction of the future \ac{los} conditions. \rev{While physical RIS placement will be limited by the environment and legal restrictions, 
RISs can still be down-selected, activated and optimized jointly for communication and localization performance. }

\subsection*{RIS Control}
\subsubsection*{Challenge}
RIS control refers to adjusting the surface impedances to steer the beams. Efficient RIS control depends on the connection to other network elements and related communication latency constraints. The material and hardware properties will set practical limits to the accuracy and speed of the phase shift control, which is in practice often quantized to finite accuracy. This may easily lead to combinatorial optimization problems. The control mechanisms and material properties have an impact on the RIS power consumption and thereby the overall system energy efficiency. All this raises research questions on how frequent the control can and should be updated (e.g., frame level or symbol level). 
\rev{This is further compounded by the possible mobility of RIS, which requires dedicated tracking routines. 
Since the RIS can operate as a transmitter, a receiver, or a reflector, each poses their own control challenges. For instance a RIS lens must control both RIS phases and switches for optimal performance. }

\subsubsection*{Opportunities}
In contrast to communication, localization and mapping applications can be supported with low  update rates, related to physical movements of the UE and environment, and hence infrequent RIS control. Each RIS with known a priori location provides an additional source of information, though RIS signals should consider multipath resolvability to avoid harmful self-interference of the controlled multipath components from the RIS. 
A priori map information, in combination with the UE location, can be leveraged to decide which RIS to activate and control, while forcing other RIS to direct signals away from the UE. Allowing limited feedback from the UE or BS helps the RIS design, e.g., phase and/or amplitude, based on predetermined codebooks at the RIS~\cite{he2019adaptive}.
The control decisions for communication will be different from those for localization, since for communication the \ac{snr} and data rate are the main metrics, while for localization accuracy and continuity are the most important.  {To this end, RISs can be controlled to optimize the \ac{gdop} or other localization-relevant metrics.} \rev{RIS control also allows reflecting an incoming signal towards multiple directions simultaneously, providing multi-user localization support from a single base station, as well independent reflections using different polarizations, frequency band, or sub-array architectures. }
Finally, the RIS activation schedule can be a tool to dim or illuminate (and thus map) parts  of the environment that are not accessible by the BS.

\subsection*{\revB{Waveform Design}}
\subsubsection*{Challenge}

The \ac{csi} or its proxy via location information needs to be used for the design of beamforming at the BS, RIS, and UE. As in standard mmWave communication, the design should be robust to account for location estimation errors, which include both position and orientation. In addition, finite quantization of the RIS phases, { which enables low-power low-complexity control as  mentioned above, adversely} limits the flexibility of the codebooks that can be used. The waveform design at the BS should also account for the presence of the RIS {and availability of \ac{los} path}~\cite{Wu2019intelligent,zheng2019intelligent}.
 
\subsubsection*{Opportunities}

\begin{figure}
 \centering
 \includegraphics[width=1\columnwidth]{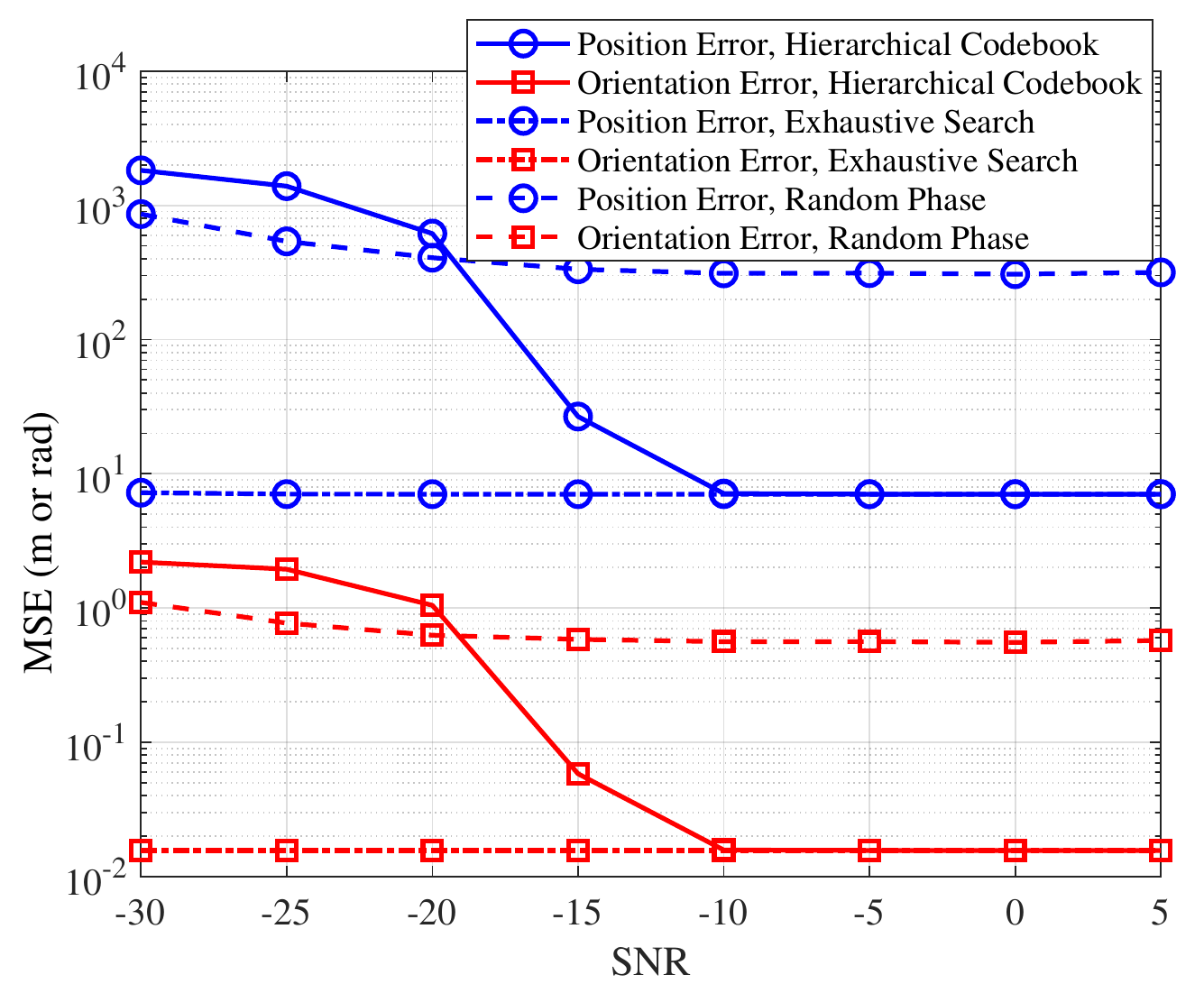}
 \caption{{Comparison on localization performance in terms of position error and orientation error with codebook-based schemes \cite{he2019adaptive}: hierarchical codebook at both UE and RIS, exhaustive search with highest-resolution codebooks at both UE and RIS, Random phase at RIS and highest-resolution codebook at UE. Parameters: BS located at $[0,0]$, RIS located at $[40, 60]$, and MS located at $[60,45]$. The number of antennas at BS and MS are 64 and 16, respectively, while the RIS has 16 units. The orientation of UE is $\pi/10$ and the \ac{los} path between BS and UE is blocked.}}
 \label{fig:PE_OE}
 \end{figure}

Similar to standard position reference signals in LTE {and 5G}, dedicated waveforms can be designed for localization with or without RIS. Such joint designs involve both waveforms at the BS as well as the codebooks at the RIS and should be sufficiently flexible to support accurate angle or delay estimation. 
The uncertainty in the map and UE location can be accounted for through robust designs  
which may explicitly encode different levels of location uncertainty. 
%
For a RIS lens transmitter, waveform design remains an unexplored area, while for a RIS reflector, preliminary results \cite{he2019adaptive} indicate the potential of dedicated designs.  Figure~\ref{fig:PE_OE} shows the performance of different codebooks at the UE and RIS, where  
a hierarchical codebook brings promising performance in terms of \ac{mse} with low training overhead and approaches that of the exhaustive search with highest resolution codebook even in the low \ac{snr} regime.
  \begin{figure*}
 \centering
 \includegraphics[width=1\textwidth]{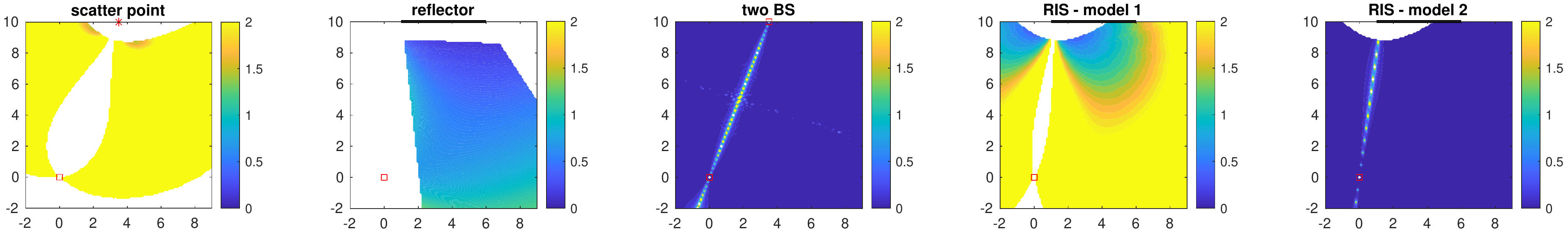}
 \caption{Comparison of far-field localization performance in terms of position error bound (shown in colors, with numbers expressed in meters) for 5 scenarios: 1 BS + 1 scatter point, 1 BS +  1 passive reflecting surface, 2 BS (each with 50\% of the bandwidth), 1 RIS with a scatter-like \revB{path loss} model per element {(model 1)}, and 1 RIS with a reflector-like \revB{path loss} model per element {(model 2)}. Parameters: BS located as $[0,0]$, 28 GHz carrier, 200 MHz bandwidth, 1 mW transmit power, RIS located at $[1.3, 10]$ with 100 elements spaced by $\lambda/2$, scatter point with RCS of $0.01~\text{m}^2$, reflecting surface with 30\% reflectivity. 
 }
 \label{fig:performance_maps}
 \end{figure*}

\subsection*{Localization and Mapping Algorithms}
\subsubsection*{Challenge}
Recovery of the user's position and the map of the environment is based on the multipath signal information. 
As signal paths parameterized by their angles and delays have no identifier of the corresponding source, this process also involves a data association of the detected paths to RIS as well as passive objects in the environment. In the presence of clutter measurements and missed detections due to directional beamforming, this is not an easy task. As the state of the user contains both the 3D position and 3D orientation, as well as a clock bias, a sufficient number of resolvable signal paths must be available, which must be enabled by proper RIS control. 
 Finally, when the user and possibly also the RIS is mobile,  dedicated tracking routines that include mobility models must be applied.

\subsubsection*{Opportunities}
The use of RIS has clear benefits for localization and mapping algorithms, since their location and orientation are known a priori. While data association is still needed to separate RIS signal paths from non-RIS signal paths, the a priori information reduces the number of data association hypotheses and allows better localization of passive landmarks and users. These benefits are present in monostatic as well as bistatic configurations, not only in terms of localization accuracy but also service coverage. 
{As for algorithm design, various solutions have been put forward for multipath-aided localization or channel-SLAM. Those could be extended to the RIS context. Among such algorithmic proposals, the solutions based on Bayesian inference over factor graphs and message-passing techniques look particularly suitable and promising, given the complexity of the new RIS-based SLAM problem (i.e., with the necessity to revolve and process signal contributions from multiple heterogeneous sources, possibly within strongly asymmetric and/or cooperative system settings)~\cite{leitinger2018belief}.}
Finally, proper algorithm design should include all aforementioned challenges in RIS localization and mapping 
to reap the full potential of the RIS.

\subsection*{Comparison of RIS and Passive Objects}
To conclude this section, we compare in Figure~\ref{fig:performance_maps} the theoretical error bounds for \ac{toa}-based localization over a canonical scene as a function of the actual UE location in five distinct scenarios:  one BS and one ``natural'' scatter  point, one BS and one passive reflecting surface, two BSs (each with $50\%$ of the bandwidth), 1 RIS with a scatter-like model per element (Model 1), and 1 RIS with a reflector-like model per element (Model 2). {Both RIS models are considered in the far field regime, for simplicity. }
{Despite the use of a single RIS in our example, it is shown that the RIS exhibiting a behavior according to Model 1 already provides limited -- yet interesting -- gains in terms of both coverage, compared to a  passive reflector, and localization error, when compared to a passive scatterer. The use of a more advanced RIS according to Model 2 could even lead to much better performance in terms of both coverage and errors, comparable to 2 active BSs.} 

\section*{Conclusions and Outlook}
{We have argued that RISs can be beneficial 
for localization and mapping in terms of improved accuracy or extended physical coverage, provided the appropriate models and algorithms can be developed. 
Progress in this area is somewhat hampered by the immaturity of working assumptions and models, which  would need further investigation and validations. Different visions of the RIS coexist today, depending on their technological maturity, leading to distinct physical behaviors (typically, in terms of end-to-end power loss over reflected paths), and, thus, distinct advantages and drawbacks with respect to localization and mapping. Beyond this, the actual feasibility of integrating and controlling the RIS at low \rev{monetary} cost, low power, low complexity and low overhead, and, possibly, the necessity to acquire side channels or prior UE location for optimal control, are still challenged by more conventional approaches such as deploying additional BSs or relays.
} 

\rev{The overall aim of this paper was to provide the reader with an up-to-date overview of RIS-based localization and mapping. We have also  described the main challenges in this field.} Moreover, we provide a large number of prominent research questions, along with potential avenues of research to answer these questions.  As we usher in the era of beyond 5G communications, we believe it is time to also consider beyond 5G or 6G localization. The RIS concept can be a game-changer for next-generation localization and mapping applications and deserves attention from the communication, signal processing, propagation, and antenna communities.




\bibliographystyle{ieeetr}
\bibliography{references}

 \balance
\begin{IEEEbiographynophoto}{Henk Wymeersch}(S'02--M'06--SM'19) received the Ph.D. degree in Electrical Engineering/Applied Sciences in 2005 from Ghent University, Belgium. He is currently a Professor in Communication Systems with the Department of Electrical Engineering at Chalmers University of Technology, Sweden. He is also a Distinguished Research Associate with Eindhoven University of Technology. Prior to joining Chalmers, he was a Postdoctoral Associate during 2006-2009 with the Laboratory for Information and Decision Systems (LIDS) at the Massachusetts Institute of Technology (MIT). He has served as Associate Editor for several IEEE Transactions.
\end{IEEEbiographynophoto}

\begin{IEEEbiographynophoto}{Jiguang He} received the B.Eng. degree from Harbin Institute of Technology, Harbin, China, in 2010, M.Sc. degree from Xiamen University, Xiamen, China, in 2013, and Ph.D. degree from University of Oulu, Finland, in 2018, all in communications engineering. From September 2013 to March 2015, he was with Key Laboratory of Millimeter Waves at City University of Hong Kong, working on beam tracking over millimeter wave MIMO systems. Since June 2015, he has been with Centre for Wireless Communications (CWC), University of Oulu, Finland. His research interests span millimeter wave MIMO communications, reconfigurable intelligent surfaces for joint communication and positioning.
\end{IEEEbiographynophoto}

\begin{IEEEbiographynophoto}{Beno\^{i}t Denis} received the E.E. (2002), M.Sc. (2002), and Ph.D. (2005) degrees in electronics and communication systems from INSA, Rennes, France. Since 2005, he has been with CEA-Leti, Grenoble, France, contributing to collaborative research projects in various domains such as wireless sensor and wearable networks, heterogeneous and cooperative networks, Ultra Wideband and (Beyond) 5G technologies, as well as mobile applications related to the Internet of Things or Intelligent Transportation Systems. His main research interests concern localization-enabled communication networks, ranging/positioning/tracking and hybrid data fusion algorithms, radio channel modeling and cross-layer protocol design.
\end{IEEEbiographynophoto}

\begin{IEEEbiographynophoto}{Antonio Clemente}received the Ph.D. degree in signal processing and telecommunications from the University of Rennes 1, Rennes, France, in 2012 at CEA-LETI, Grenoble, France. Since 2013, he is a Research Engineer at CEA-LETI, Grenoble, France. His current research interests include fixed-beam and electronically reconfigurable transmitarray antennas, millimeter-wave and sub-THz antennas, antenna arrays, near-field focused systems, antenna modelling, miniature integrated antennas, antenna fundamental limitations, near-field and far-field antenna measurements. He has authored or co-authored more than 105 papers in international journals and conferences, and received 12 patents. From 2013 to 2016, he has been the technical coordinator of the H2020 joint Europe and South Korea 5GCHAMPION project.
\end{IEEEbiographynophoto}

\begin{IEEEbiographynophoto}{Markku Juntti} (S'93--M'98--SM'04--F'20) received his Dr.Sc.\ (EE) degree from University of Oulu, Oulu, Finland in 1997. In the academic year 1994--95, he was a Visiting Scholar at Rice University, Houston, Texas. In 1999--2000, he was a Senior Specialist with Nokia Networks. Dr.\ Juntti has been a professor of communications engineering since 2000 at University of Oulu, Centre for Wireless Communications (CWC), where he also serves as Head of CWC -- Radio Technologies Research Unit. His research interests include signal processing for wireless networks as well as communication and information theory. Dr.\ Juntti is also an Adjunct Professor at Rice University. He is an Editor of \textsc{IEEE Transactions on Communications}.
\end{IEEEbiographynophoto}

\end{document}